\documentclass[aps,prl,reprint,groupedaddress]{revtex4-1} 

\usepackage{graphicx}
\usepackage{amsmath}
\usepackage{amssymb}

\begin{document}


\title{Quasiparticle Damping of Surface Waves in Superfluid $^3$He and $^4$He}


\author{M.~S.~Manninen}%
\email{msmannin@boojum.hut.fi}%
\author{J. Rysti}%
\author{I. Todoshchenko}%
\author{J. Tuoriniemi}%
\affiliation{O.V. Lounasmaa Laboratory, Aalto University, P.O. Box 15100, 00076 Aalto, Finland}


\date{\today}

\begin{abstract}
Oscillations on free surface of superfluids at the inviscid limit are damped by quasiparticle scattering. 
We have studied this effect in both superfluids $^3$He and $^4$He deep below the respective critical temperatures. 
Surface oscillators offer several benefits over immersed mechanical oscillators traditionally used for similar purposes.
Damping is modeled as specular scattering of ballistic quasiparticles from the moving free surface. 
The model is in reasonable agreement with our measurements for superfluid $^4$He but significant deviation is found for $^3$He.
\end{abstract}


\maketitle


All sorts of oscillating bodies have been used for long to study dissipation mechanisms in superfluids~\cite{%
Andronikashvilli1946,
Einzel1987,
Morishita1989,
Fisher1989,
Zadorozhko2009,
Bradley2009,
Jager1995,
Gonzalez2013
}. 
The purpose is often to determine the density of thermal quasiparticles in order to deduce the temperature of the superfluid. 
The influence of the quasiparticles is to damp the motion of the oscillator as the quasiparticles 
bounce off the surfaces carrying away fraction of the momentum in the event. 
Theoretical treatment of the process requires the knowledge of the roughness of the surfaces 
leading to either 
specular~\cite{Morishita1989,Fisher1991} or diffuse~\cite{Enrico1995} scattering, 
or more generally some intermediate of the two extremes.
Also, any solid object has much larger density as compared to helium, so that relative change in momentum per event is very small, 
thus limiting the sensitivity of practical devices.
Moreover, mechanical oscillators suffer from internal damping of the device itself.

Free surface of superfluid helium set into oscillatory motion is probably the most ideal tool for studying 
the interaction of quasiparticles with impenetrable boundaries. 
No additional mass besides helium itself is involved in the motion and the quasiparticle scattering is presumably perfectly specular.
This has been experimentally verified in the case of superfluid $^4$He~\cite{Baddar96}.
As the temperature is reduced deep below the superfluid transition temperature $T_c$, 
the quasiparticle mean free path increases very rapidly and at about $T\approx T_c/4$ it can usually be assumed 
to exceed the typical dimensions of the experiment. 
This means that the quasiparticles essentially do not interact with each other, 
just with the boundaries of the fluid volume, and can be treated as ballistic entities. 
This simplifies the theoretical treatment a great deal.

The crucial difference between the two helium isotopes is their bosonic ($^4$He) or fermionic ($^3$He) character, 
which largely dictates their behavior at very low temperatures. 
Whereas bosonic (even number of elementary particles) $^4$He is superfluid below 2~K,
fermionic (odd number of elementary particles) $^3$He atoms need to first form pairs leading to rather 
complex superfluid properties but only at temperatures a thousand times lower than in $^4$He~\cite{VollhardtWolfle1990}. 
At saturated vapor pressure (practically zero pressure) $^3$He becomes superfluid at about 1~mK and exhibits an isotropic B-phase, 
unless sufficiently large magnetic field is applied. 

Different quantum statistics of the two helium isotopes implies that the thermal quasiparticles have entirely different character 
in these two superfluids and their dispersion relation and scattering properties differ in fundamental ways. 
Therefore, it is of interest to study both of these media in the very same experimental cell 
with no alterations in the geometry using oscillations of the free surface as the indicator of the quasiparticle properties.

Surface wave resonances in helium fluids have been utilized for determining the surface tension 
but only in $^4$He have such studies been extended to the superfluid state~\cite{Iino85b}. 
However, very little data is available on the temperature dependence of damping of such resonances. 
We are aware of only one systematic study of the damping of the surface waves in $^4$He, 
utilizing electrons trapped on the free surface~\cite{Sommerfeld96}. 
This work unfortunately suffered from technical difficulties preventing from drawing any firm conclusions on the topic. 
In bulk superfluid $^3$He only one prior report on observation of surface waves exists~\cite{Eltsov13arxiv}. 
Thin superfluid films support another type of oscillation, labeled as third sound, 
which has been observed on both $^4$He~\cite{Roche95} and $^3$He~\cite{Schechter98}.

In this article we present a simple model for the damping of the surface waves on bulk superfluid helium in the ballistic 
quasiparticle limit and compare the results with our measurements on both $^4$He and $^3$He in the exact same geometry.


\begin{figure}%
\includegraphics{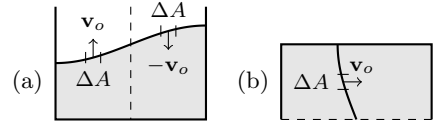}%
\caption{\label{fig:analog}Alternative formulation of a standing surface wave (a), %
where the two halves can be redrawn with a common impervious interface (b) moving with velocity $\mathbf{v}_o$.}%
\end{figure}%
Let us consider specular scattering of ballistic quasiparticles from a moving object, free surface in particular.
Indeed, surface waves can be modeled as a moving object in fluid from the quasiparticle point of view
since for any surface element with area $\Delta A$ moving with velocity $\mathbf{v}_{o}$
one finds a counterpart, another surface element moving with the same speed but to opposite direction $-\mathbf{v}_{o}$,
see Fig.~\ref{fig:analog}.

According to the momentum and energy conservation laws the energy difference between
incoming ($E_{1}$) and scattered ($E_{2}$) excitations is
\begin{equation}
\Delta E = E_{2}-E_{1} = \mathbf{v}_{o} \cdot \left( \mathbf{p}_{2} - \mathbf{p}_{1} \right)
= \mathbf{v}_{o} \cdot \Delta \mathbf{p}
\text{,}
\label{eq:conservationlaw}
\end{equation}
where $\mathbf{p}_{1}$ and $\mathbf{p}_{2}$ are the momenta of the incoming and scattered excitations respectively.

\begin{figure}%
\includegraphics{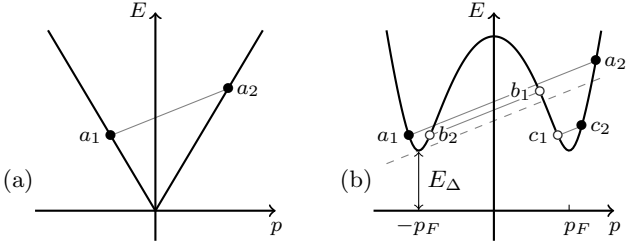}%
\caption{\label{fig:quasiparticlesSMALL}%
Energy spectra of collective excitations in 1D for bosonic $^4$He (a) and fermionic $^3$He (b) in the superfluid state. %
In $^4$He the spectrum at low energies is linear (phononic branch) %
and the excitations are particle-like ($\bullet$) quasiparticles. %
In $^3$He the relevant excitations are so-called Bogoliubov quasiparticles with isotropic energy gap $E_\Delta$ %
at Fermi momentum $p_F$ and there are both particle-like ($\bullet$) and hole-like ($\circ$) quasiparticles. %
Representative scattering processes $a$, $b$, and $c$ from the moving surface are marked with gray lines and explained in the text. %
The slope of the lines correspond to the velocity of the moving surface, which is exaggerated for clarity in the figure. %
Fermi quasiparticles below the dashed line in (b) experience Andreev reflection and contribute very little to the momentum transfer.}%
\end{figure}%

Damping of the moving object is due to the elastic momentum transfer $\Delta \mathbf{p}$ between the object and quasiparticles.
The damping force $\mathbf{F}$ is
\begin{equation}
\mathbf{F}=\int_\Omega
\frac{
   -\Delta \mathbf{p} \left|\left( \mathbf{v}_g -\mathbf{v}_{o}\right)\cdot\mathbf{\hat{v}}_{o} \right| \Delta A
   }{
   h^3\left\{\exp[E/(k_B T)] \pm 1\right\}
   }
d\mathbf{p}
\text{,}
\label{eq:damping}
\end{equation}
where
$\mathbf{v}_g=\nabla_{\mathbf{p}} E$ is the group velocity of quasiparticles with energy $E$,
$\left|\left( \mathbf{v}_g -\mathbf{v}_{o}\right)\cdot\mathbf{\hat{v}}_{o} \right|\Delta A$
is their volumetric flow rate towards the moving object,
and $d\mathbf{p}/\textbf{(}h^{3} \{\exp[E/(k_B T)] \pm 1\}\textbf{)}$
is the number density of thermally excited quasiparticles within $d\mathbf{p}$ for bosons ($-$) or fermions ($+$) at temperature $T$.

In bosonic $^4$He the low-energy quasiparticles can be considered as phonon-like excitations with
energy $E=u\left|\mathbf{p}\right|$
and constant group velocity,
$\mathbf{v}_g= u \mathbf{\hat{p}}$
as sketched in Fig.\,\ref{fig:quasiparticlesSMALL}a in 1D.
Since $\mathbf{v}_g \uparrow \uparrow \mathbf{p}$ they are always particle-like excitations.
According to Eqs.~(\ref{eq:conservationlaw}) and (\ref{eq:damping}) the damping force in 3D is
\begin{equation}
\mathbf{F}_{B}=-\frac{8\pi^5 (k_B T)^4 }{15 h^3 u^4}  \Delta A \mathbf{v}_{o}
=-P_{B}(T)\Delta A \mathbf{v}_{o}
\text{,}
\label{eq:damping4He}
\end{equation}
which defines the temperature dependent factor $P_B(T)$ for bosonic excitations.
The only medium dependent parameter here is the speed of sound $u$, 
which is $u=239~\text{m/s}$ in $^4$He at saturated vapor pressure in the zero temperature limit~\cite{Abraham1970}.

In fermionic $^3$He the Bogoliubov quasiparticle energy spectrum is more complicated,
see Fig.\,\ref{fig:quasiparticlesSMALL}b.
As for superconductors
there is a BCS energy gap $E_\Delta=1.764~k_B T_c$ at Fermi momentum $p_F=8.28{\times}10^{-25}~\text{kg\,m/s}$~\cite{Dobbs2000}.
In addition, besides the particle-like quasiparticles there are also hole-like quasiparticles where 
$\mathbf{v}_g \uparrow \downarrow \mathbf{p}$.
As depicted in Fig.\,\ref{fig:quasiparticlesSMALL}b the normal particle-to-particle or hole-to-hole scattering
is not always allowed but a quasiparticle
may be Andreev-reflected from hole $c_1$ to particle $c_2$ (or vice versa) in the process
transferring only negligible amount of momentum compared to normal scattering~\cite{Fisher1989}.
Thus Eq.~(\ref{eq:damping}) gives
\begin{equation}
\mathbf{F}_{F}=-\frac{4\pi p_F^4}{h^3\exp\left[E_\Delta/(k_B T)\right]} \Delta A \mathbf{v}_{o}
=-P_{F}(T)\Delta A \mathbf{v}_{o}
\label{eq:damping3He}
\end{equation}
in the limit $\left|\mathbf{v}_{o}\right| p_F \ll k_B T \ll k_B T_c$,
where the Andreev-scattered states ($E<E_\Delta+2\mathbf{p}\cdot\mathbf{v}_{o}$) with negligible momentum transfer 
are excluded from the region of integration $\Omega$
allowing to use $\Delta\mathbf{p}=2p_F (\mathbf{\hat{p}}\cdot\mathbf{\hat{v}}_{o})  \mathbf{\hat{v}}_{o}$ within $\Omega$.

For an arbitrary standing surface wave mode in any geometry oscillating at frequency $f$,
the vertical deviation from the equilibrium can be written as
$z=z_0(x,y)\cos(2\pi f t)$.
The total energy of the wave is
\begin{equation}
E_\text{total} = \frac{1}{2}\rho g \int_A z_0^2   \,dA
\text{,}
\label{eq:TotalEnergy}
\end{equation}
where $\rho$ and $g$ are the fluid density and gravitational acceleration, respectively, while 
the surface energy has been neglected here.
This is legit for wavelengths much longer than the capillary length of the fluid under inspection. 
In helium fluids this is a safe assumption below frequencies of about 20~Hz.

According to Eqs.~(\ref{eq:damping4He}) and (\ref{eq:damping3He}) the damping force is
proportional to the area and the velocity of the free surface element with temperature dependent multiplicative factor $-P(T)$,
different for bosons and fermions.
The total energy dissipated in one cycle is
\begin{equation}
E_\text{loss} = \int_A\int_{t=0}^{1/f} P\mathbf{v}_{o}^2 \,dt \,dA
= 2\pi^2 f P \int_A z_0^2   \,dA
\text{.}
\label{eq:dissipatedEnergy}
\end{equation}
The quality factor $Q$ of the oscillation is then
\begin{equation}
Q=2\pi\frac{E_\text{total}}{E_\text{loss}}=\frac{\rho g}{2\pi f P}=\frac{f}{\Delta f}
~\,\Rightarrow\,~
\frac{\Delta f}{f^2}=\frac{2\pi P(T)}{\rho g}
\text{,}
\label{eq:Qfactor}
\end{equation}
where $\Delta f$ is the resonance frequency width.
An important observation is that the scaled frequency width $\Delta f /f^2 $ does not depend on the geometry of the surface 
nor on the resonance frequency but only on temperature and known physical parameters.


In our experiment helium was refrigerated by a nuclear demagnetization cryostat \cite{Yao2000} in
a cell with a central cuboid volume (length 10~mm, width 10~mm, height 25~mm) that is connected from 
two opposite corners to a surrounding annular channel (diameter 25~mm, width 1~mm, height 25~mm). 
The bottom of the central cuboid was 0.4~mm lower than that of the annulus.
A photograph of the cell can be found in Ref.~\onlinecite{Manninen2013}.

The surface level and its local oscillations were detected capacitively with two independent interdigital capacitors 
mounted on opposite vertical walls of the cuboid volume. 
The surface waves could be generated either by ambient vibrational noise or by an active drive~\cite{Manninen2013}.

Temperature readings are most reliable above 10~mK or so, where several independent thermometers were available in the cryostat 
and good thermal contact with the helium sample can be guaranteed. 
Another firm calibration point was provided by the superfluid transition temperature of $^3$He,
while deep in the superfluid state of $^3$He we had to rely on adiabatic changes of magnetic field on the copper nuclear refrigerant.
At the lowest temperatures below 0.2~mK inevitable large thermal gradients between the helium sample and the copper refrigerant
developed despite of extensive sintered heat exchangers on all available surfaces of the sample cell.


\begin{figure}%
\includegraphics{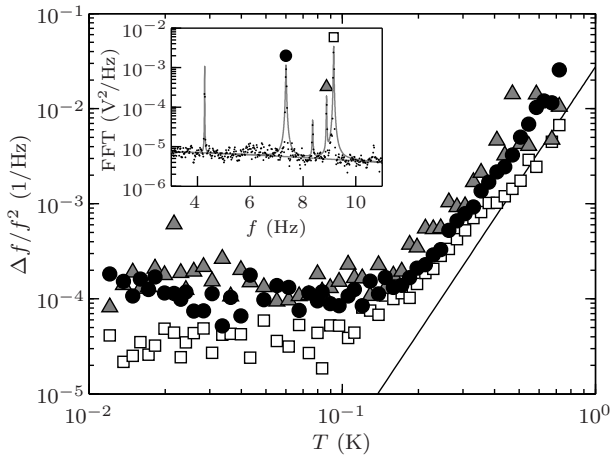}%
\caption{\label{fig:Super4He_width}Resonance frequency width $\Delta f$ %
scaled by the resonance frequency squared $f^2$ in superfluid $^4$He. %
The solid line represents the expected behavior according to Eqs.~(\ref{eq:damping4He}) and (\ref{eq:Qfactor}). %
There are no fitted parameters. %
The inset shows an example of the frequency spectra as driven by ambient noise at $T=15~\text{mK}$ %
together with a fitted curve. %
The peaks indicated by the three symbols refer to the data in the main frame. %
The finite frequency resolution results in leveling off of the data to about $\Delta f\approx10~\text{mHz}$ %
at the lowest temperatures.}%
\end{figure}%
Surface wave resonances in superfluid $^4$He were observed up to 60~Hz frequency, including several higher order modes.
Scaled frequency widths $\Delta f/f^2$ of the clearest resonances are shown in Fig.~\ref{fig:Super4He_width} 
with an example of raw frequency spectra below 11~Hz in the inset. 
At higher frequencies surface tension would become more and more significant altering the simple frequency scaling used.
Depth of the helium pool was $h=5.2~\text{mm}$ in the cuboid volume in this case.

The resonance frequencies of the low frequency modes do not exactly match those expected from the geometry and, in fact, 
there are more peaks visible in our data than the rectangular and annular surfaces should support. 
As explained above, though, it is not necessary to know the geometry and whereabouts of the wave, 
once we scale the resonance width by the resonance frequency squared. 
This takes care of complications due to possibly poorly defined geometry. 
It is reassuring that the data so scaled fall into roughly unified set and turn closely to the theoretical curve 
with no fitting parameters whatsoever. 
At the lowest temperatures the resonances become so narrow that there is essentially just one spectrum point off the baseline 
at each resonance,
limiting the lowest measurable frequency width to about 10~mHz.
As the width is expected to scale as $f^2 T^4$, the modes with the lowest frequencies are deemed to lose resolution to 
width at higher temperature than the modes with higher frequencies.

\begin{figure}%
\includegraphics{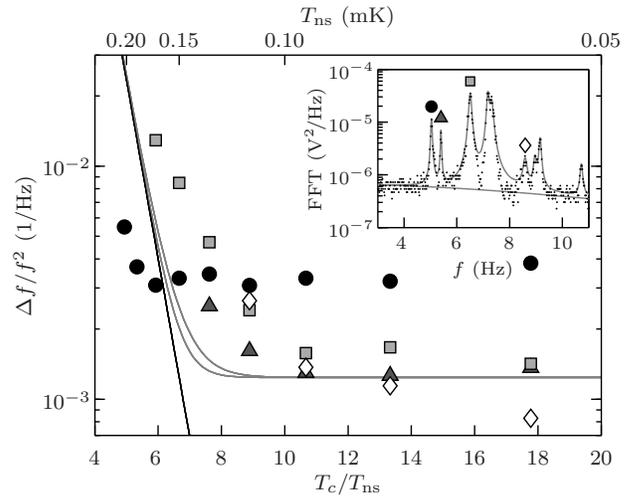}%
\caption{\label{fig:Super3He_width}Scaled resonance frequency width $\Delta f/f^2$ in superfluid $^3$He %
as a function of inverse temperature of the nuclear stage $T_\text{ns}$, %
scaled by the superfluid transition temperature $T_c\approx1~\text{mK}$. %
The solid straight line is the expected behavior according to Eqs.~(\ref{eq:damping3He}) and (\ref{eq:Qfactor}). %
An example of raw spectra with a fit is shown in the inset at $T_\text{ns}=0.11~\text{mK}$. %
The resonance width remains within resolution at any temperature, %
as $^3$He cannot be cooled as deep into the superfluid state as $^4$He in relative terms. %
Instead, the leveling off is obviously due to loosing thermal contact to the sample at about $T_c/T\approx$~7---10. %
The then expected behavior~\cite{Parpia1985} is illustrated by the lower of the bending curves. %
The other one assumes a constant temperature independent extra damping %
of $\Delta f/f^2=1.2\times10^{-3}~\text{Hz}^{-1}$.}%
\end{figure}%
In superfluid $^3$He the surface wave resonances were observable below about $0.2~T_c$.
Scaled frequency widths of selected resonances are shown in Fig.~\ref{fig:Super3He_width}
with a representative frequency spectrum in the inset.
Here the helium depth was $h=3.9~\text{mm}$ in the cuboid volume.
The peaks excluded from the analysis seemed to consist of multiple resonances inseparable from one other. 
This could be deduced on the basis of driven resonances, an example of which is shown in Ref.~\onlinecite{Manninen2013}.

There are remarkable deviations from the expected behavior
but the correspondence cannot be improved by treating the energy gap $E_\Delta$ as a free parameter.
The best resolved resonance (square symbols in Fig.~\ref{fig:Super3He_width}) 
and two other sparser sets of data (triangles and diamonds) form more or less consistent set,
though the exponential temperature dependence is much weaker than expected on the basis of 
Eqs.~(\ref{eq:damping3He}) and (\ref{eq:Qfactor}). 
Yet another set (circles, the lowest frequency resonance) is completely bizarre: 
the resonance remains quite broad at the lowest temperatures, 
which could be explained by somewhat higher temperature in the annular volume where that mode most likely resides, 
but then it becomes \textit{too narrow} at highest temperatures, 
while the conditions most definitely should have become better equilibrated in terms of temperature
no matter where that particular mode was located. 
We emphasize that this anomaly is not due to any analysis artifact, 
as this tendency is clearly visible in the raw spectra as well.

The leveling off of the width at the lowest temperatures is not due to inadequate spectral resolution in this case, 
as the setup was exactly the same as that for $^4$He capable of resolving ten times narrower resonances. 
Instead, helium temperature probably just saturated due to Kapitza resistance, 
which was treated according to an empirical model~\cite{Parpia1985}
to produce the lower of the saturating curves in Fig.~\ref{fig:Super3He_width}. 
It does not reproduce the data particularly well, 
mainly because the temperature dependent part has too shallow temperature dependence. 
Alternatively, some other additional dissipation mechanism could be assumed besides ballistic quasiparticles, 
which would result temperature independent contribution to $P_F(T)$.
This results the upper bending curve in Fig.~\ref{fig:Super3He_width},
which is not much better and could not be distinguished from the Kapitza effect given the statistics of our data. 
%
%
For the time being we are not able to fully explain these results on $^3$He.

Setting aside the problem of not perfectly fitting the theory, we can still comment on the sensitivity to temperature 
of the surface wave resonances deep in the superfluid state. 
All other types of oscillators utilized so far lose their sensitivity because the damping due to the fluid practically 
vanishes and the device displays merely its own internal damping at the lowest possible temperatures in superfluid helium. 
This is not so for the surface wave resonator in $^3$He. 
There was still a margin of about a factor ten in the present experiment before the instrumental resolution would have become 
the limiting factor, and there probably is room for improving that somewhat, too. 
If we interpret the leveling off of the data to be caused by the saturating temperature, 
we get the lowest helium temperature as about $T_c/T\approx 10$, 
which is roughly the same as the lowest temperatures measured in $^3$He ever~\cite{Bradley2004}.


In conclusion, the measured damping of surface waves in superfluid $^4$He between 0.1---0.6~K corresponds well with the model 
of specular scattering of ballistic quasiparticles from the oscillating free surface. 
The low temperature limit of sensitivity was set by the instrumental resolution cutting off the temperature dependence 
below about 100---150~mK.
In superfluid $^3$He, however, there is a remarkable discrepancy between the specular scattering model and the experiment. 
In this case the resolution was not limited by the measuring scheme but, instead, 
either by additional damping mechanisms in superfluid helium or 
by the saturating temperature not following that of the refrigerator at the very lowest temperatures.
Any adjustment of the energy gap value suggested by the BCS theory does not improve the correspondence between our data 
and the theory.
\begin{acknowledgments}
This work was supported in part by European Union (FP7/2007-2013, Grant No.~228464 Microkelvin) and 
by Academy of Finland (LTQ CoE grant no.~250280). 
This research made use of the Aalto University Low Temperature Laboratory infrastructure. 
We also acknowledge the grants from Jenny and Antti Wihuri Foundation, and the National Doctoral Programme in Materials Physics.
We thank V.~B.~Eltsov, P.~J.~Heikkinen, J.-P.~Kaikkonen, A.~J.~Niskanen, A.~Ya.~Parshin, V.~Peri, 
A.~Salmela, A.~Sebedash and V.~Tsepelin
for valuable discussions and assistance.
\end{acknowledgments}


\bibliography{Name}

\end{document}